\documentstyle[preprint,prc,aps,epsfig]{revtex}
\begin{document}

\catcode`\@=11
\def\eqalign#1{\null\,\vcenter{\openup\jot\m@th
\ialign{\strut\hfil$\displaystyle{##}$&$\displaystyle{{}##}$\hfil
     \crcr#1\crcr}}\,}
\catcode`\@=12

\title{NN Scattering Phase Shifts in a Chiral Constituent Quark Model}
\author{D. Bartz\footnote{e-mail: d.bartz@ulg.ac.be} and Fl. Stancu
\footnote{e-mail: fstancu@ulg.ac.be}}
\address{Universit\'{e} de Li\`ege, Institut de Physique B.5, Sart Tilman,
B-4000 Li\`ege 1, Belgium}
\date{\today}
\maketitle

\begin{abstract}
We study the nucleon-nucleon interaction within a chiral constituent quark
model which reproduces succesfully the baryon spectra. We calculate the $^3S_1$
and $^1S_0$ phase shifts by using the resonating group method. They clearly
indicate the presence of a strong repulsive interaction at short distance,
due to the spin-flavour symmetry of the quark-quark interaction and of the
quark interchange between the two interacting nucleons. A $\sigma$-meson
exchange
quark-quark interaction, providing a medium-range attraction, helps to get
closer to the experimental phase shifts.
\end{abstract}

\vspace{0.7cm}
PACS number(s): 24.85.+p, 21.30.-x, 13.75.Cs

\vspace{0.5cm}

\begin{flushleft}
{\bf Introduction}
\end{flushleft}

The aim of the present work is to study the nucleon-nucleon (NN) interaction and in particular the NN scattering phase shifts in the framework of a chiral constituent quark model which has proved succesfull in baryon spectroscopy \cite{GR96,GPP96,GPP97,GPVW98,Gl00}. In the calculation of the NN phase shifts we employ the resonating group method (RGM) \cite{Wh37,Ka77} which has been first applied to nuclear physics. This method is very convenient for treating the interaction between composite particles. In the NN problem each nucleon is supposed to be composed of three quarks and the interaction between quarks is here provided by the same chiral quark model, both for quarks belonging to the same nucleon or to different nucleons.

The study of the NN interaction in the framework of quark models has already some history. We restrict our discussion to nonrelativistic models where the RGM can be applied. Twenty years ago Oka and Yazaki \cite{OY80} published the first nucleon-nucleon $L=0$ phase shifts calculated with the resonating group method. The work \cite{OY80} and subsequent developments are revised in \cite{We85}. Those results were obtained from models based on one-gluon exchange (OGE) interaction between quarks. Based on such models one could explain the short-range repulsion of the NN interaction potential as due to the chromomagnetic spin-spin interaction, combined with quark interchanges between $3q$ clusters.  In order to describe the data, long- and medium-range interactions were added at the nucleon level.

Here we employ a constituent quark model where the short-range quark-quark interaction is entirely due to pseudoscalar meson exchange instead of one-gluon exchange. This is the chiral constituent quark model proposed in Ref. \cite{GR96} and parametrized in a nonrelativistic version in Refs. \cite{GPP96,GPP97}. A semirelativistic version is also presented in Ref. \cite{GPVW98}. The present status of this model is summarized in Ref. \cite{Gl00}. The spin-flavour symmetry structure of the model is getting support from the phenomenological analysis of 
$L=1$ negative parity baryon resonances \cite{CG99}. Also lattice \cite{Li99} and $1/N_C$ QCD studies \cite{Ca00} have a consistent interpretation in a constituent quark model with pseudoscalar meson exchange interaction.

The origin of the model \cite{GR96,GPP96,GPP97,GPVW98,Gl00} is thought to lie in the spontaneous breaking of chiral symmetry in QCD which implies the existence of constituent quarks with a dynamical mass and Goldstone bosons (pseudoscalar mesons). According to the two-scale picture of Manohar and Georgi \cite{MG94}, at a distance beyond that of spontaneous chiral symmetry breaking, but within that of the confinement scale, the appropriate degrees of freedom should be the constituent quarks and the chiral meson fields. If a quark-pseudoscalar meson coupling is assumed, in a nonrelativistic limit one obtains a quark-meson vertex proportional to $\vec{\sigma} \cdot \vec{q}\ \lambda^F$ with $\vec{\sigma}$ the Pauli matrices, $\vec{q}$ the momentum of the meson and $\lambda^F$ the Gell-Mann flavour matrices. This generates a pseudoscalar meson exchange interaction between quarks which is spin and flavour dependent. In the following, the model \cite{GR96,GPP96,GPP97,GPVW98,Gl00}, based entirely on this interaction, is referred to as the Goldstone boson exchange (GBE) model.

It is important to correctly describe both the baryon spectra and the baryon-baryon interaction with the same model. The model \cite{GR96,GPP96,GPP97,GPVW98,Gl00} gives a good description of the baryon spectra and in particular the correct order of positive and negative parity states, both in nonstrange and strange baryons, in contrast to the OGE model. In fact the pseudoscalar exchange interaction has two parts : a repulsive Yukawa potential tail and an attractive contact $\delta$-interaction. When regularized \cite{GPP96,GPP97,GPVW98}, the latter generates the short-range part of the quark-quark interaction. This dominates over the Yukawa part in the description of baryon spectra. The whole interaction contains the main ingredients required in the calculation of the NN potential, and it is thus natural to study the NN problem within the GBE model. In addition, the two-meson exchange interaction between constituent quarks reinforces the effect of the flavour-spin part of the one-meson exchange and also provides a contribution of a $\sigma$-meson exchange type \cite{RB99} required to describe the middle-range attraction.

Preliminary studies with this interaction have been made in Refs. \cite{SPG97,BS99a,BS99b}. This work is a natural extension of these previous studies. Ref. \cite{SPG97} was rather exploratory about the role of a spin-flavour dependent interaction in giving rise to a repulsive core. Within the parametrization \cite{GPP96} of the GBE model it was found that at zero-separation between two $3q$ clusters the height of the repulsive core is 0.830 GeV and 1.356 GeV in the $^3S_1$ and $^1S_0$ channels respectively. The spin-flavour symmetry and the parametrization \cite{GPP96} of the GBE model favours the $|[42]_O[51]_{FS}\rangle$ state, which becomes highly dominant. In Ref. \cite{BS99a}, instead of cluster model states, a better basis obtained from single-particle molecular type states, has been employed. The situation has been found to be similar, the repulsion being reduced by about 200 MeV in the $^3S_1$ channel and by about 400 MeV in the $^1S_0$ channel. This is natural because the molecular orbital basis gives a lower bound of the expectation value of the Hamiltonian in the six-quark basis. In Ref. \cite{BS99b} an adiabatic NN potential was calculated based on the model \cite{GPP96}. It was found that none of the bases, cluster or molecular, leads to an attractive pocket. An attraction was simulated by introducing a $\sigma$-meson exchange of an analytic form similar to that of the pseudoscalar meson exchange of \cite{GPP96}. 

To our knowledge, the present study is the first derivation of NN scattering phase shifts with the full SU(3) version of the GBE model which provides a comparison with the experimental data. Similar RGM calculations \cite{SG00}, where a simplified SU(2) version of the GBE model has been used, have also shown the presence of a short-range repulsion in the behaviour of the phase shifts in the $L = 0$ channel. However that paper did not aim at a comparison with the experiment.
\begin{flushleft}
{\bf The model}
\end{flushleft}

The GBE Hamiltonian considered below has the form \cite{GPP97}
\begin{equation}\label{HAMILTONIAN}
H=\sum_i m_i+\sum_{i=1}{\frac{p^2_i}{2m_i}}-K_G+\sum_{i<j} V_{Conf}(r_{ij})+\sum_{i<j} V_{\chi}(r_{ij})\ ,
\end{equation}
where $K_G$ is the kinetic energy of the center of mass. The linear confining interaction is
\begin{equation}
V_{Conf}(r_{ij})=-\frac{3}{8}\lambda^c_i \cdot \lambda^c_j~ (C r_{ij} + V_0)
\end{equation}
and the spin-spin component of the GBE interaction in its $SU_F(3)$ form is
\begin{equation}
V_{\chi}(r_{ij})=\{ {\sum_{F=1}^3 V_{\pi}(r_{ij}) \lambda^F_i \lambda^F_j + \sum_{F=4}^7 V_K(r_{ij}) \lambda^F_i \lambda^F_j + V_{\eta}(r_{ij}) \lambda^8_i \lambda^8_j + \frac{2}{3} V_{\eta'}(r_{ij}) } \} \vec{\sigma_i} \cdot \vec{\sigma_j}\ .
\end{equation}
The interaction (3) contains $\gamma=\pi,K,\eta,$ and $\eta'$ meson-exchange terms and $V_{\gamma}(r_{ij})$ is given as the sum of two distinct contributions: a Yukawa-type potential containing the mass of the exchanged meson and a short-range contribution of opposite sign, the role of which is crucial in baryon spectroscopy. For a given meson $\gamma$, the exchange potential is
\begin{equation}
V_{\gamma}(r) = \frac{g_{\gamma}^2}{4\pi} \frac{1}{12m_im_j} \{ \mu_{\gamma}^2 \frac{e^{- \mu_{\gamma} r}}{r} - \Lambda_{\gamma}^2 \frac{e^{- \Lambda_{\gamma} r}}{r} \} \ ,
\end{equation}
where $\Lambda_{\gamma}=\Lambda_0+\kappa \mu_{\gamma}$. For a system of $u$ and $d$ quarks only, as is the case here, the $K$ exchange does not contribute. In the calculations below we use the parameters of Refs. \cite{GPP97}. These are
$$m_{u,d}=340\ {\rm MeV},\ \ C=0.77\ {\rm fm}^{-2},$$
$$\mu_{\pi}=139\ {\rm MeV},\ \mu_{\eta}=547\ {\rm MeV},\ \mu_{\eta'}=958\ {\rm MeV},$$
$$\frac{g_{\pi q}^2}{4\pi} = \frac{g_{\eta q}^2}{4\pi} = 1.24,\ \ \frac{g_{\eta' q}^2}{4\pi} = 2.7652,$$
\begin{equation}
\ \ \ \Lambda_0=5.82\ {\rm fm}^{-1},\  \kappa = 1.34,\  V_0=-112\ {\rm MeV}\ .
\end{equation}

The reason of using the parametrization \cite{GPP97}, instead of \cite{GPP96}, as in the previous work \cite{SPG97,BS99a,BS99b}, is that it is more realistic. Its volume integral, i.e. its Fourier transform at $\vec{q} = 0$, vanishes, consistently with the quark-pseudoscalar meson vertex proportional to $\vec{\sigma} \cdot \vec{q}\ \lambda^F$. In addition this interaction does not enhance the quark-quark matrix elements containing $1p$ relative motion, as it is the case with the parametrization \cite{GPP96}. This point has been raised in Ref. \cite{SG99}. As a net result, in the parametrization \cite{GPP96} the attraction due to the Yukawa-potential tail is overwhelmed by the repulsion resulting from the short-range part of the hyperfine interaction \cite{BS99b}.

Some more comments are in order before proceeding further. 1) The above parametrization gives a good description of baryon spectra. We do not change any parameter obtained from the fit \cite{GPP97}. Such a parametrization is, of course, only effective.  However, irrespective of the parametrization, the flavour-spin symmetry is essential in this model. 2) There are also semirelativistic versions of the GBE model, as for example \cite{GPVW98} but first one has to study whether or not the resonating group method is applicable to semirelativistic six-quark Hamiltonians. The present RGM approach \cite{Ka77} lies heavily on an $s^3$ structure of the nucleon wave function and such a simple description is inadequate for semirelativistic wave functions. Anyhow, for a fair comparison with results based so far on nonrelativistic OGE models a nonrelativistic version of the GBE model is desirable.
\newpage

\begin{flushleft}
{\bf The nucleon}
\end{flushleft}

In the RGM approach the wave function of the ground state nucleon must be known (see next section). As indicated above, its orbital part $\phi$ has an $s^3$ structure.  This is a very good approximation to the exact wave function. The function $\phi$ is fully symmetric with respect to any permutation of $S_3$ and is chosen of the form
\begin{equation}\label{NUCLEONGROUND}
\phi=\prod_{i=1}^3 g(\vec{r}_i,b)\ ,
\end{equation}
with $g(\vec{r}_i,b)$ given by
\begin{equation}\label{QUARKGAUSS}
g(\vec{r},b)=(\frac{1}{\pi b^2})^{3/4} e^{-\frac{r^2}{2 b^2}}\ ,
\end{equation}
The size parameter $b$ appearing in (\ref{QUARKGAUSS}) is obtained variationally from the stability condition (see for example Ref.\cite{OY80})
\begin{equation}\label{STABILITY}
\frac{\partial}{\partial b}\langle \phi|H|\phi\rangle =0\ ,
\end{equation}
where $H$ is the Hamiltonian (\ref{HAMILTONIAN}) written for a $3q$ system. This condition gives $b=0.44$ fm, which we use below. 

\begin{flushleft}
{\bf The six-quark state}
\end{flushleft}

From symmetry considerations (see for example \cite{Ha81}) we can find which channels contribute to a totally antisymmetric six-quark state of a given spin $S$ and isospin $I$ when the orbital, spin, flavour and colour degrees of freedom of the nucleon are taken into account. For example for $SI = (10)$ or $(01)$ the channels are NN, $\Delta\Delta$ and the hidden colour CC. In this work the discussion is restricted to NN channels only. Preliminary studies with coupled channels indicate that the influence of $\Delta\Delta$ and CC channels on the phase shifts is very small \cite{BSfuture}.

\newpage
\begin{flushleft}
{\bf The resonating group method}
\end{flushleft}

Here we shortly describe the resonating group method. A detailed account of its application to the NN scatering can be found in the original papers by Oka and Yazaki \cite{OY80}. The resonating group method is a well established method for studying the interaction between two composite particles. First it has been succesfully applied to the derivation nucleus-nucleus interaction (see e.g. \cite{Ka77}). Since the work of Oka and Yazaki, it is being used to calculate NN phase shifts assuming a quark structure for the nucleons. In the NN system each nucleon is treated as a $3q$ cluster. In the one channel approximation of the RGM the NN system is described by a $6q$ wave function of the form
\begin{equation}
\Phi(\vec{\xi}_A,\vec{\xi}_B,\vec{R}_{AB})={\cal A}[\phi(\vec{\xi}_A)\phi(\vec{\xi}_B)\chi(\vec{R}_{AB})]\ ,
\end{equation}
where $\vec{\xi}_A=(\vec{\xi}_1,\vec{\xi}_2)$ and $\vec{\xi}_B=(\vec{\xi}_3,\vec{\xi}_4)$ are the internal coordinates of the clusters $A$ \& $B$ and $\vec{R}_{AB} = \vec{R}_{A} -\vec{R}_{B}$ is the relative coordinate between the two clusters. Thus $\phi_i\ (i=A,B)$ are the internal wave functions of the clusters which are supposed to be known and $\chi$ is the ${\it unknown}$ wave function describing their relative motion. The development below is based on the assumption that $\phi_i$ has a simple $s^3$ structure which is reasonable for nonrelativistic models. The antisymmetrization operator $\cal A$ is
\begin{equation}
{\cal A} = 1 - \sum_{i=1}^3 \sum_{j=4}^6 P_{ij}\ .
\end{equation}
From the variational principle one can obtain the equation determining the relative wave function $\chi(\vec{R}_{AB})$ as
\begin{equation}\label{VAR}
\int \phi^+(\vec{\xi}_A)\phi^+(\vec{\xi}_B)(H-E)\Phi(\vec{\xi}_A,\vec{\xi}_B,\vec{R}_{AB})d^3\xi_Ad^3\xi_B = 0\ ,
\end{equation}
where $H$ is the Hamiltonian (\ref{HAMILTONIAN}) of the six-quark system. We introduce the Hamiltonian kernel
\begin{equation}
{\cal H}(\vec{R},\vec{R'})=\int \phi^+(\vec{\xi}_A)\phi^+(\vec{\xi}_B) \delta(\vec{R}-\vec{R}_{AB}) H {\cal A} [\phi(\vec{\xi}_A)\phi(\vec{\xi}_B) \delta(\vec{R'}-\vec{R}_{AB})] d^3\xi_A d^3\xi_B d^3R_{AB}
\end{equation}
and the normalization kernel
\begin{equation}
{\cal N}(\vec{R},\vec{R'})=\int \phi^+(\vec{\xi}_A)\phi^+(\vec{\xi}_B) \delta(\vec{R}-\vec{R}_{AB}) {\cal A} [\phi(\vec{\xi}_A)\phi(\vec{\xi}_B) \delta(\vec{R'}-\vec{R}_{AB})] d^3\xi_A d^3\xi_B d^3R_{AB}\ .
\end{equation}
Then Eq. (\ref{VAR}) can be written as
\begin{equation}
\int {\cal L}(\vec{R},\vec{R'}) \chi(\vec{R'}) d^3R' = 0\ ,
\end{equation}
where ${\cal L}(\vec{R},\vec{R'}) = {\cal H}(\vec{R},\vec{R'}) - E {\cal N}(\vec{R},\vec{R'})$. This is the RGM equation. Here it has been solved by using the method of Ref. \cite{Ka77}. Accordingly, the relative wave function $\chi (\vec{R})$ has been expanded over a finite number of equally displaced Gaussians. For scattering states this expansion holds up to a finite distance $R=R_c$, where  $R_c$ depends on the range of the interaction. Beyond $R_c$, $\chi(\vec{R})$ is written as the usual combination of Hankel functions containing the $S$-matrix. Then the phase shifts are determined by imposing the continuity of  $\chi(\vec{R})$ and of its derivative with respect to $R$ at  $R=R_c$. For the NN channel a number of 15 Gaussians is large enough to obtain convergence and the matching radius is $R_c = 4.5$ fm. The size parameter of the Gaussians is fixed at $b=0.44$ fm, as discussed above. An important analytic step in solving the Eq. (11) is the calculation of two-body matrix elements contained in the Hamiltonian kernel ${\cal H}(\vec{R},\vec{R'})$ and the normalization kernel  ${\cal N}(\vec{R},\vec{R'})$. All relevant two-body matrix elements obtained by integration in the color, spin and flavor spaces are given in Table 1. Symbolically an upper index $f$ is introduced in order to distinguish between the color $\lambda_i^c$ and the flavor $\lambda_i^f$ $SU(3)$ matrices. 

\begin{flushleft}
{\bf Results}
\end{flushleft}

In Figs. 1 and 2 we plot the calculated $^3S_1$ and $^1S_0$ as a function of $E_{lab}$. For a comparison, we add the corresponding results of \cite{FF82} obtained with the spin-spin part of the one-gluon exchange (OGE) model. The phase shifts of both models reveal the presence of a short-range repulsion in the NN interaction. In the GBE model, without the long-range Yukawa part \cite{BSfuture}, the repulsion is stronger and corresponds to a hard core radius $r_0^{GBE} = 0.81$ fm (versus $r_0^{OGE} = 0.35$ fm) for $^1S_0$ and $r_0^{GBE} = 0.68$ fm (versus $r_0^{OGE} = 0.30$ fm) for $^3S_1$. This outcome is consistent with the findings of Ref. \cite{SG00} where a simplified $SU(2)$ version of the GBE model has been used. Thus the GBE model can explain the short-range repulsion, as due to a flavour-spin quark-quark interaction and to the quark interchange between clusters. However, to describe the scattering data and the deuteron properties, intermediate- and long-range attraction potentials are necesary. In calculations based on OGE models they were phenomenologically simulated at the NN level by central and tensor potentials respectively \cite{OY83} (see also Ref. \cite{We85}). However, for a consistent picture it is desirable to search for the origin of the attraction at the quark level. Here we incorporate a $\sigma$-meson exchange interaction at the quark level and study its effect on the $^1S_0$ phase shift. Note that the $^1S_0$ phase shift is not influenced by a tensor potential. For the $\sigma$-meson exchange interaction we choose the following form
\begin{equation}\label{SIGMA}
V_{\sigma}=-\frac{g_{\sigma q}^2}{4\pi}~(\frac{e^{-\mu_{\sigma}r}}{r}-\frac{e^{-\Lambda_{\sigma}r}}{r})\ ,
\end{equation}
with parameters discussed below. The introduction of such an interaction is consistent with the spirit of the GBE model. It simulates the effect of two correlated pions \cite{RB99}. The good quality of the baryon spectrum is not distroyed by the addition of this interaction which essentially leads to an overall shift of the spectrum \cite{SS00}. The sensitivity of the $^1S_0$ phase shift with respect to the coupling constant $\frac{g_{\sigma q}^2}{4\pi}$, the mass $\mu_{\sigma}$ and the cut-off parameter $\Lambda_{\sigma}$ can be seen from Figs. 3-5. As expected, the attraction in the NN potential increases with $\frac{g_{\sigma q}^2}{4\pi}$ and hence the value of $E_{lab}$ where the phase shift changes sign also increases with $\frac{g_{\sigma q}^2}{4\pi}$, as shown in Fig. 3. Note that the potential $V_{\sigma}$ of (\ref{SIGMA}) remains attractive as long as $\mu_{\sigma} < \Lambda_{\sigma}$. However $\mu_{\sigma}$ cannot be too close to $\Lambda_{\sigma}$. As suggested by Fig. 4, the attractive pocket in the NN potential becomes too small for $\mu_{\sigma} > $ 0.65 GeV, making the repulsion dominant and leading to negative phase shifts at all energies, when $\frac{g_{\sigma q}^2}{4\pi}$ = 1.24 and $\Lambda_{\sigma}$ = 0.830 GeV.
A large difference between  $\mu_{\sigma}$ and $\Lambda_{\sigma}$ is not good either. From  Fig. 5 one can see that when $\Lambda_{\sigma} > $ 0.95 GeV and  $\mu_{\sigma}$ = 0.6 GeV
an undesired bound state in the $^1S_0$ phase shift is accomodated at $\frac{g_{\sigma q}^2}{4\pi}$ = 1.24. This is due to the fact that the contribution of the second term in the right hand side of (\ref{SIGMA}) becomes negligible and $V_{\sigma}$ brings too much attraction in the NN potential.

In this way we found  an optimal set of values
\begin{equation}
\frac{g_{\sigma q}^2}{4\pi} = \frac{g_{\pi q}^2}{4\pi} = 1.24,~~~~~
\mu_{\sigma} = 0.600\ {\rm GeV}\ ,~~~~~\Lambda_{\sigma} = 0.830\ {\rm GeV}\ .
\end{equation}
  
As one can see from Fig. \ref{Fig. 6}, with these values the theoretical curve gets quite close to the experimental points without altering the good short-range behaviour, and in particular the change of sign of the phase shift at $E_{lab} \approx 260$ MeV. Thus the addition of a $\sigma$-meson exchange interaction alone leads to a good description of the phase shifts in a large energy interval. One can argue that the still existing discrepancy at low energies could possibly be removed by the coupling of the $^5D_0$ N-$\Delta$ channel, suggested by Ref. \cite{VF95} in the frame of a hybrid model, containing both gluon and meson exchange at the quark level. To achieve this coupling, as well as to describe the $^3S_1$ phase shift, the introduction of a tensor interaction is necessary.
 
\begin{flushleft}
{\bf Conclusion}
\end{flushleft}

This study shows that the GBE model gives good promise in describing the NN scattering properties, while offering a good description of baryon spectra. In the following step it will be interesting to include a tensor interaction at the quark level, which does not alter the quality of the spectrum \cite{Wa00} and which provides the long-range attraction in the NN potential.

\begin{flushleft}
{\bf Acknowledgements}
\end{flushleft}

We are most grateful to Kiyotaka Shimizu for introducing us into the intricancies of the resonating group method and for help in checking some partial results.


\begin{figure}
\begin{center}
\psfig{figure=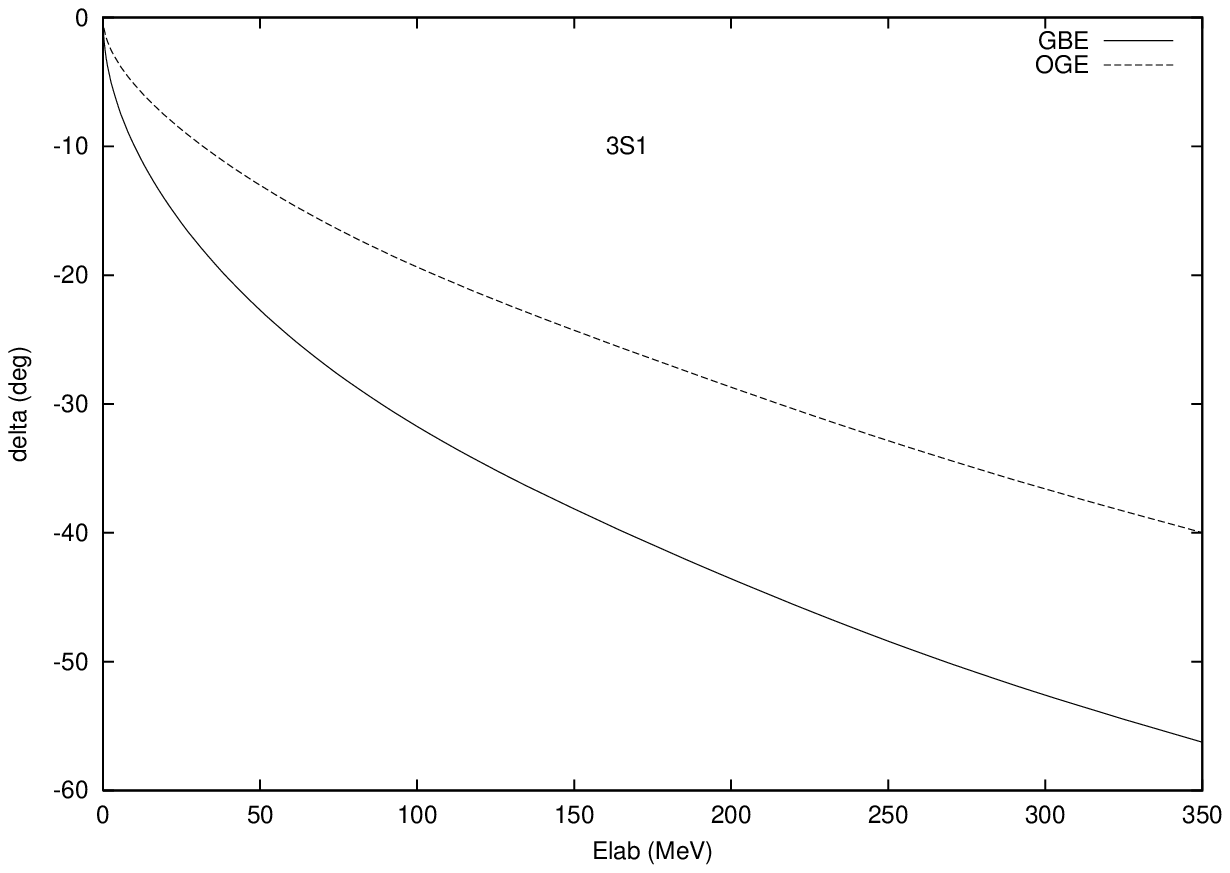}
\end{center}
\caption{\label{Fig. 1} The $^3S_1$ NN scattering phase shift as a function of
$E_{lab}$. The solid line shows the result obtained in the GBE model, the
dashed line in the OGE model of Ref. [19]. }
\end{figure}

\begin{figure}
\begin{center}
\psfig{figure=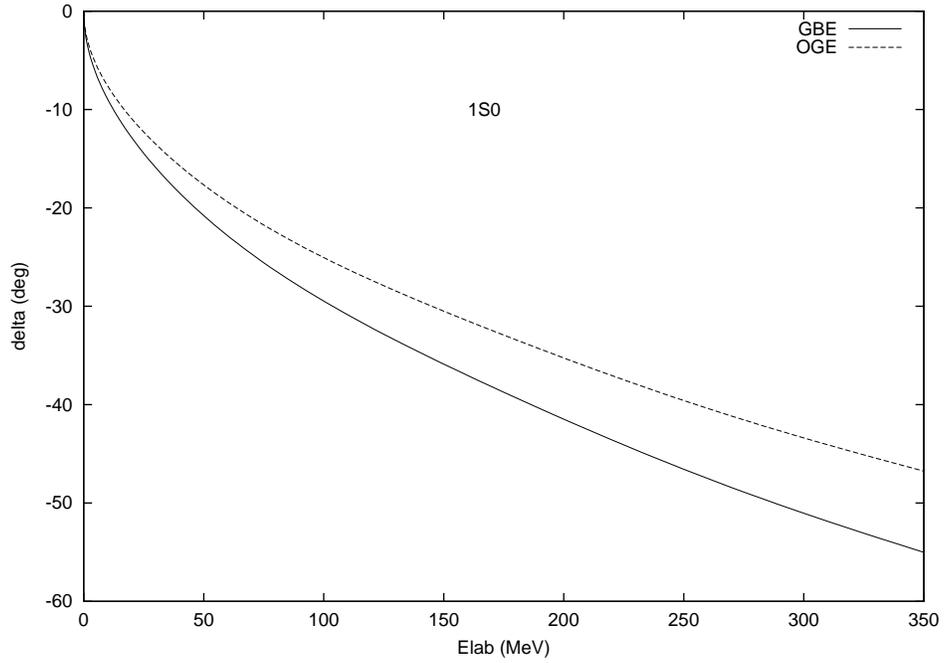}
\end{center}
\caption{\label{Fig. 2} Same as Fig. 1 but for the $^1S_0$ NN partial wave.}
\end{figure}

\begin{figure}
\begin{center}
\psfig{figure=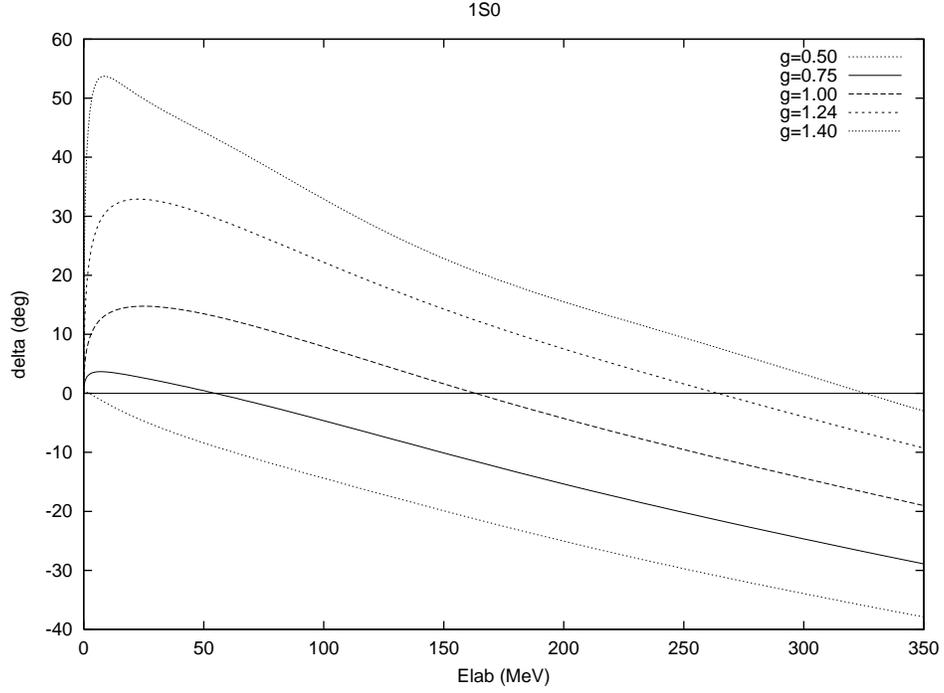}
\end{center}
\caption{\label{Fig. 3} The $^1S_0$ phase shift for various  
values of $g = g_{\sigma q}^2/4\pi$ at fixed $\mu_{\sigma}$ = 0.600 GeV
and $\Lambda_{\sigma}$ = 0.830 GeV.}
\end{figure}

\begin{figure}
\begin{center}
\psfig{figure=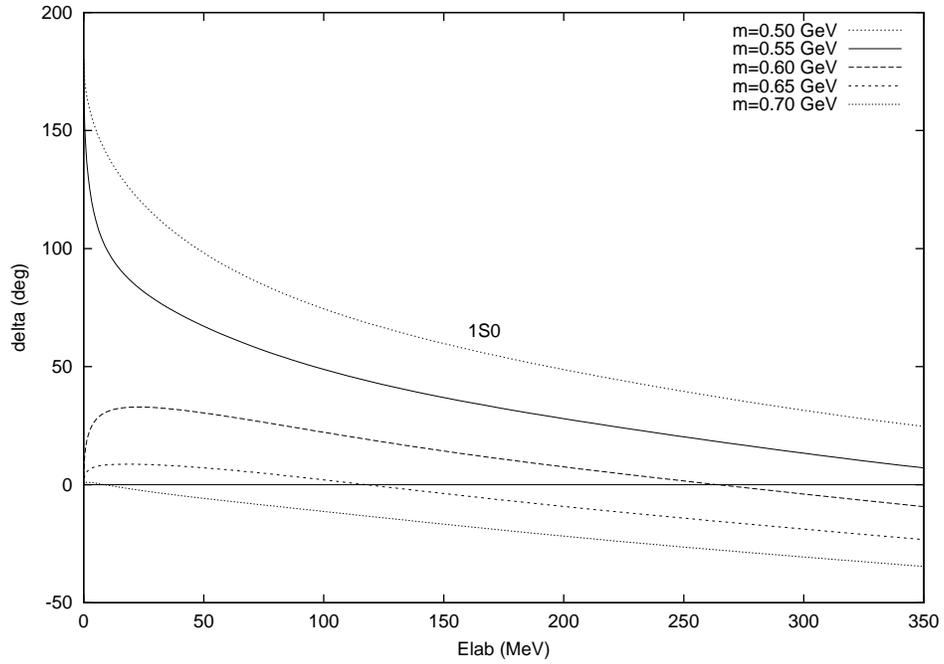}
\end{center}
\caption{\label{Fig. 4} The $^1S_0$ phase shift for various 
values of $m = \mu_{\sigma}$ at fixed  $g_{\sigma q}^2/4\pi$ = 1.24
and $\Lambda_{\sigma}$ = 0.830 GeV.}
\end{figure}

\begin{figure}
\begin{center}
\psfig{figure=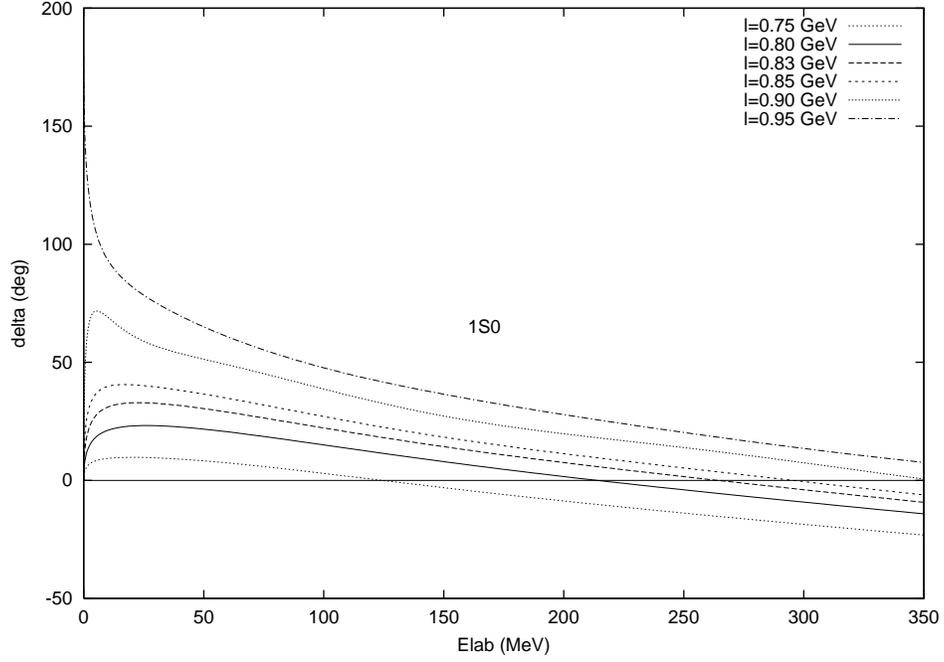}
\end{center}
\caption{\label{Fig. 5} The $^1S_0$ phase shift for various
values of $l = \Lambda_{\sigma}$ at fixed  $g_{\sigma q}^2/4\pi$ = 1.24
and $\mu_{\sigma}$ = 0.600 GeV.}
\end{figure}

\begin{figure}
\begin{center}
\psfig{figure=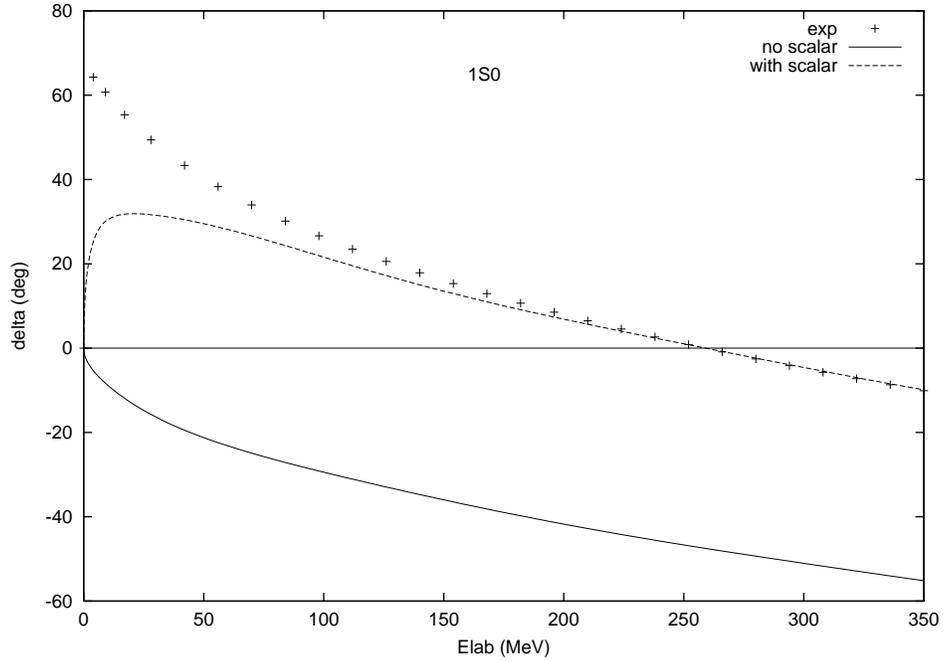}
\end{center}
\caption{\label{Fig. 6} The $^1S_0$ NN scattering phase shift obtained in the
GBE model as a function of $E_{lab}$. The solid line is the same as that of
Fig. 1 and the dashed line includes the effect of the $\sigma$-meson exchange
potential (12) between quarks with $\mu_{\sigma}=0.600$ GeV and
$\Lambda_{\sigma}=0.830$ GeV. Experimental data are from Ref. [25].}
\end{figure}

\begin{table}
\renewcommand{\arraystretch}{0.65}
\parbox{18cm}{\caption[matrixelements]{\label{matrixelements}  Matrix elements
$\langle NN|O|NN\rangle$ of different operators $O$ for $(S,T)$ = $(1,0)$ and
$(0,1)$.}}
\begin{tabular}{ccc}
$O$ & $(S,T)=(1,0)$ & $(S,T)=(0,1)$ \\
\tableline
$1$                                                                       
&  243 &  243 \\
$P_{36}^{f \sigma c}$                                                  
&   -3 &   -3 \\
$\lambda_1^c \cdot \lambda_2^c$                                                 
& -648 & -648 \\
$\lambda_3^c \cdot \lambda_6^c$                                                 
&    0 &    0 \\
$\lambda_1^c \cdot \lambda_2^c\ P_{36}^{f \sigma c}$                            
&    8 &    8 \\
$\lambda_3^c \cdot \lambda_6^c\ P_{36}^{f \sigma c}$                            
&  -16 &  -16 \\
$\lambda_1^c \cdot \lambda_3^c\ P_{36}^{f \sigma c}$                            
&    8 &    8 \\
$\lambda_1^c \cdot \lambda_6^c\ P_{36}^{f \sigma c}$                            
&    8 &    8 \\
$\lambda_1^c \cdot \lambda_4^c\ P_{36}^{f \sigma c}$                            
&   -4 &   -4 \\
$\sigma_1 \cdot \sigma_2$                      
& -243 & -243 \\
$\sigma_3 \cdot \sigma_6$                      
&   27 &  -81 \\
$\sigma_1 \cdot \sigma_2 \ P_{36}^{f \sigma c}$ 
&   51 &   51 \\
$\sigma_3 \cdot \sigma_6 \ P_{36}^{f \sigma c}$ 
&   57 &   93 \\
$\sigma_1 \cdot \sigma_3 \ P_{36}^{f \sigma c}$ 
&  -21 &  -21 \\
$\sigma_1 \cdot \sigma_6 \ P_{36}^{f \sigma c}$ 
&  -21 &  -21 \\
$\sigma_1 \cdot \sigma_4 \ P_{36}^{f \sigma c}$ 
&    6 &    0 \\
$\sigma_1 \cdot \sigma_2\ \tau_1 \cdot \tau_2$                                        
& 1215 & 1215 \\
$\sigma_3 \cdot \sigma_6\ \tau_3 \cdot \tau_6$                                        
& -225 & -225 \\
$\sigma_1 \cdot \sigma_2\ \tau_1 \cdot \tau_2\ P_{36}^{f \sigma c}$                   
& -111 & -111 \\
$\sigma_3 \cdot \sigma_6\ \tau_3 \cdot \tau_6\ P_{36}^{f \sigma c}$                   
&  177 &  177 \\
$\sigma_1 \cdot \sigma_3\ \tau_1 \cdot \tau_3\ P_{36}^{f \sigma c}$                   
&   33 &   33 \\
$\sigma_1 \cdot \sigma_6\ \tau_1 \cdot \tau_6\ P_{36}^{f \sigma c}$                   
&   33 &   33 \\
$\sigma_1 \cdot \sigma_4\ \tau_1 \cdot \tau_4\ P_{36}^{f \sigma c}$                   
&    9 &    9 \\
$\sigma_1 \cdot \sigma_2\ \lambda_1^{f} \cdot \lambda_2^{f}$                          
& 1134 & 1134 \\
$\sigma_3 \cdot \sigma_6\ \lambda_3^{f} \cdot \lambda_6^{f}$                          
& -216 & -252 \\
$\sigma_1 \cdot \sigma_2\ \lambda_1^{f} \cdot \lambda_2^{f}\ P_{36}^{f \sigma c}$     
&  -94 &  -94 \\
$\sigma_3 \cdot \sigma_6\ \lambda_3^{f} \cdot \lambda_6^{f}\ P_{36}^{f \sigma c}$     
&  196 &  208 \\
$\sigma_1 \cdot \sigma_3\ \lambda_1^{f} \cdot \lambda_3^{f}\ P_{36}^{f \sigma c}$     
&   26 &   26 \\
$\sigma_1 \cdot \sigma_6\ \lambda_1^{f} \cdot \lambda_6^{f}\ P_{36}^{f \sigma c}$     
&   26 &   26 \\
$\sigma_1 \cdot \sigma_4\ \lambda_1^{f} \cdot \lambda_4^{f}\ P_{36}^{f \sigma c}$     
&   11 &    9 \\
\tableline
factor & $\frac{1}{243}$ & $\frac{1}{243}$ \\
\end{tabular}
\end{table}

\end{document}